\documentclass[a4paper]{jpconf}
\usepackage{color}
\usepackage{graphicx}
\begin{document}
\title{Recent results of gluon and sea quark polarization measurements in polarized proton-proton collisions at STAR}

\author{Xuan Li$^{[1]}$ for the STAR Collaboration}

\address{1) Physics Department, Temple University, 1900 North 13th street, Philadelphia, 19122, USA}

\ead{xuanli@rcf.rhic.bnl.gov}

\begin{abstract}
The STAR experiment at RHIC is carrying out a comprehensive high-energy spin physics program to understand the internal structure and dynamics of the proton in polarized proton-proton collisions at $\sqrt{s} = 200$ GeV and $\sqrt{s} = 500/510$ GeV. STAR has the capability, with nearly full azimuthal coverage, to reconstruct leptons, hadrons and jets in the mid-rapidity region ($|\eta|<1$). The results for inclusive jet longitudinal double spin asymmetries taken during the 2009 RHIC run indicate the first non-zero gluon contribution ($\Delta g(x,Q^{2}) / g(x,Q^{2})$) to the proton spin for $0.05<x<1$ (Bjorken-x: momentum fraction of partons). Recent longitudinal single-spin asymmetry measurements of $W^{+/-}$ bosons at $\sqrt{s} = 500/510$ GeV in polarized proton-proton collisions provide a direct probe of the polarized anti-u and anti-d quark distributions ($\Delta \bar{u}(x,Q^{2})$, $\Delta \bar{d}(x,Q^{2})$). These results better constrain the polarized gluon and sea quark distributions of the proton in the RHIC sensitive kinematic region. Future measurements with continuing high energy polarized proton-proton run at RHIC and detector upgrade will explore the gluonic contribution to the proton spin in extended range.
\end{abstract}

\section{Introduction}
%$\frac{\hbar}{2}$
The discovery of the so-called ``proton spin crisis" \cite{EMC_1988} has prompted a variety of investigations to determine precisely the various constituent spin-contributions. The sum of the quark and anti-quark polarization contributions ($\Delta\Sigma$) measured in polarized deep inelastic scattering (DIS) experiments is less than one third of the total proton spin \cite{dis, dis1}. The limited Bjorken-x region accessed by the fixed target polarized DIS experiments provides poor constraints on the polarized gluon distribution $\Delta g(x,Q^{2})$ and flavor separated polarized anti-quark distributions $\Delta \bar{q}(x,Q^{2})$ . The helicity structure of the proton is being studied at the Relativistic Heavy Ion Collider (RHIC) via longitudinally polarized proton-proton collisions. RHIC is a versatile machine, capable of colliding longitudinally and tranversely polarized protons at varying center of mass energies as well as unpolarized heavy ions at different energies and atomic numbers. Since 2009, RHIC has collided longitudinally polarized proton beams at $\sqrt{s} = 200$ GeV and $\sqrt{s} = 500/510$ GeV with improved luminosity by a factor of over 100 relative to previous years and polarization over $50\%$. The Solenoidal Tracker at RHIC (STAR) has both tracking detectors and electromagnetic calorimeters with nearly full azimuthal coverage, therefore final states like jets and electrons can be well reconstructed, especially at mid-rapidity ($|\eta|<1$). 

Jets can be treated as surrogates of the hard scattering partons. STAR has measured inclusive jet and di-jet production in proton-proton collisions. A global QCD fit which includes the 2009 STAR inclusive jet double spin asymmetry $A_{LL}$ in 200 GeV p+p collisions, extracts a non-zero gluon spin contributions ($\Delta g(x,Q^{2}) / g(x,Q^{2})$) to the proton in $0.05<x<1$ \cite{2009_jet_All, DSSV_new}. W production in polarized proton-proton collisions which avoids the parton fragmentation processes is a cleaner probe than hadrons to detect the light sea quark $\bar{u}$ and $\bar{d}$ polarization contribution ($\Delta \bar{u}(x,Q^{2}) / \bar{u}(x,Q^{2})$, $\Delta \bar{d}(x,Q^{2}) / \bar{d}(x,Q^{2})$) to the proton. The first lepton rapidity dependent $W^{+/-}$ single spin asymmetry $A_{L}$ in 500/510 GeV proton-proton collisions indicates a sizable and positive $\bar{u}$ quark polarization \cite{2012_W_AL_paper}. The latest released inclusive jet and W spin results at STAR promote our understanding of the gluon, $\bar{u}$ and $\bar{d}$ polarized distributions in the probed $0.05<x<1$ region. Details about the jet and W spin analyses and projections for the future measurments will be discussed in the following sections.

\begin{figure}[t]
\begin{minipage}{17pc}
\includegraphics[width=17pc]{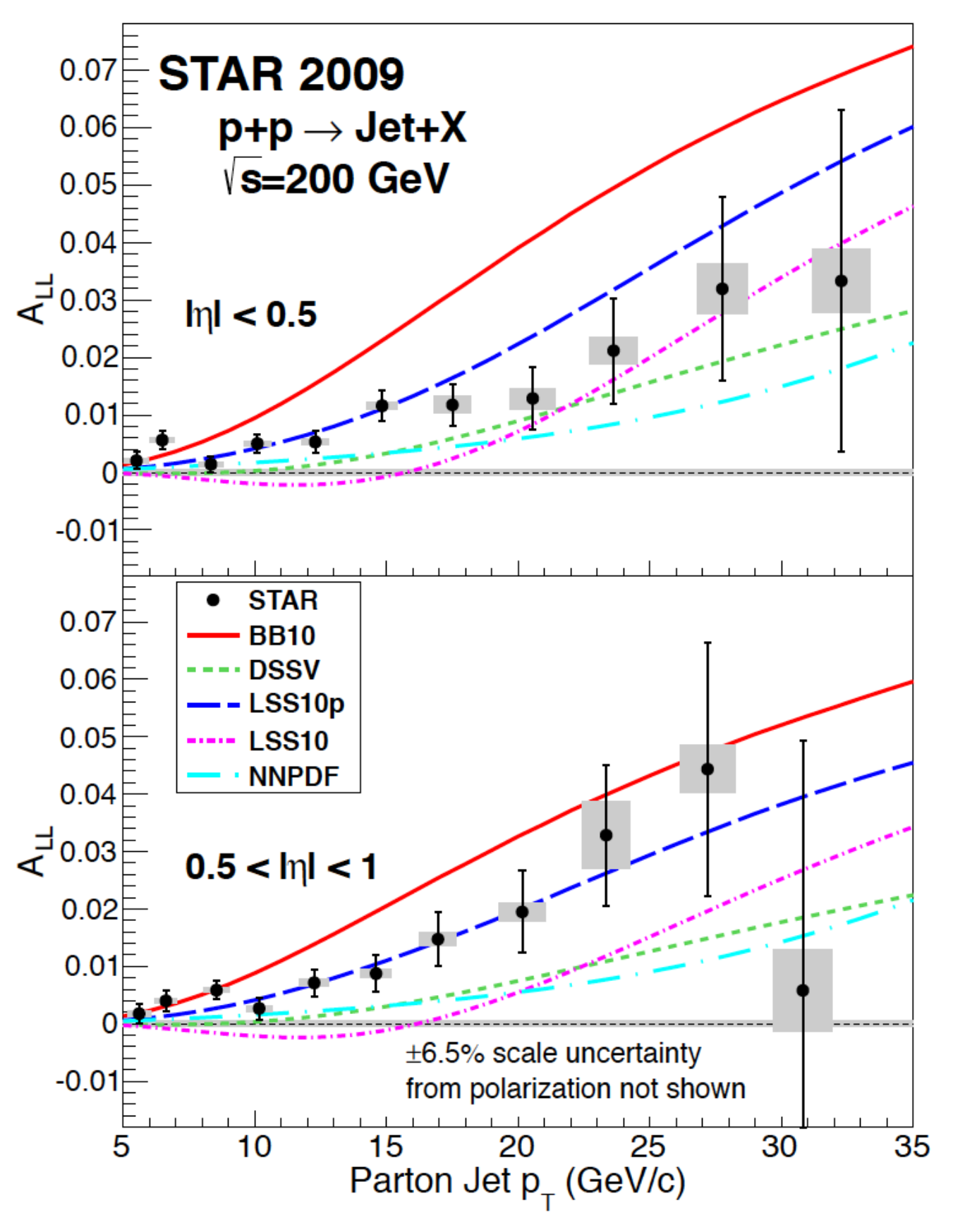} %\hspace{0.5pc}
\caption{\label{2009jet_plot} \it STAR 2009 inclusive jet $A_{LL}$ versus parton jet $p_{T}$ in 200 \textup{GeV} p+p collisions \cite{2009_jet_All}. $A_{LL}$ of inclusive jets with $|\eta|<0.5$ is shown in the top panel, and the results with $0.5<|\eta|<1.0$ are shown in the bottom panel. Error bars are statistical and gray boxes stand for systematic errors.}
\end{minipage} 
\hfill\begin{minipage}{16pc}
\includegraphics[width=16pc]{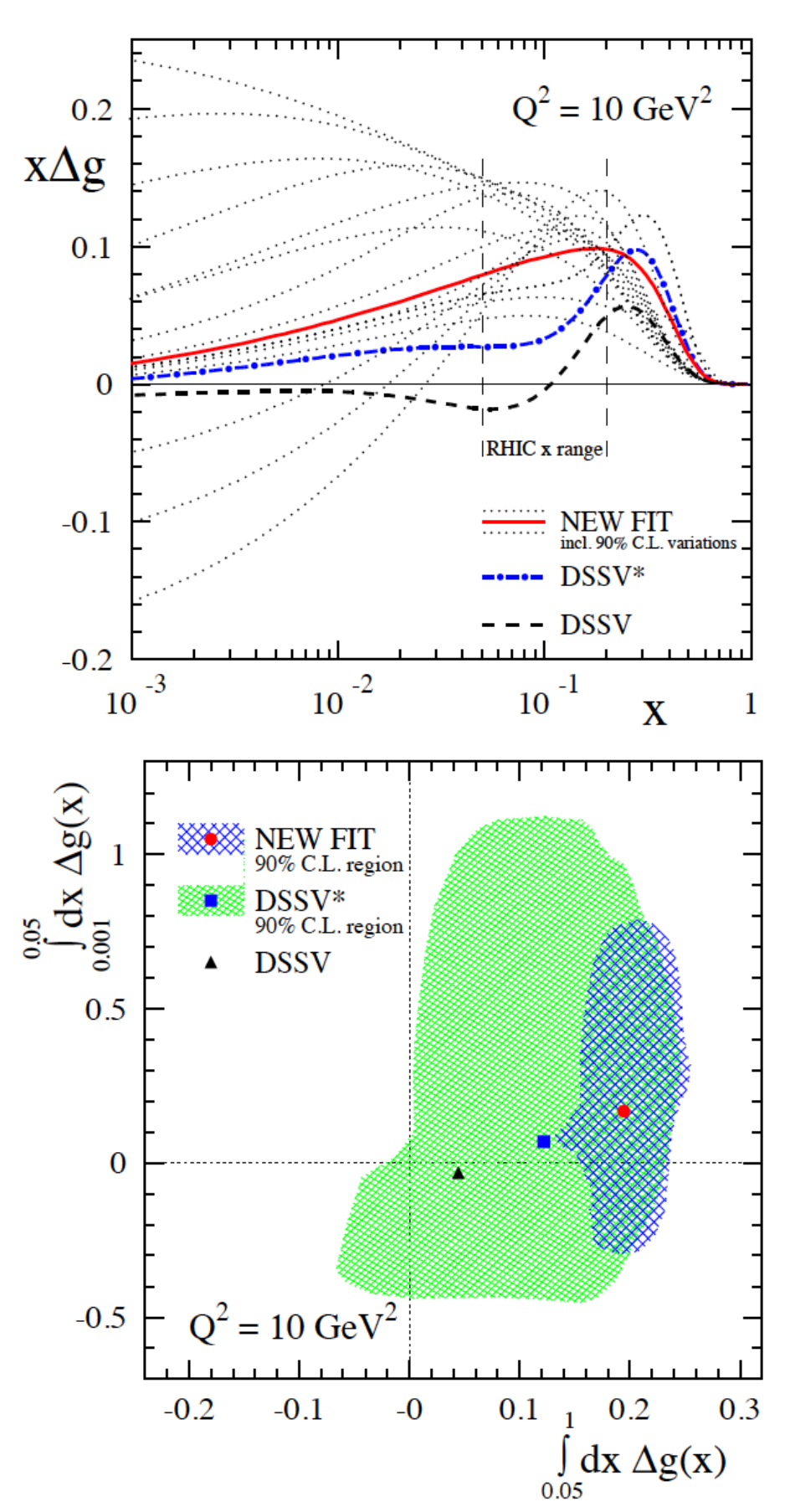}
\caption{\label{DSSV_Dg} \it Recent global analysis from the DSSV group. Top panel: the gluon helicity distribution $\Delta g$ at $Q^{2} = 10$  \textup{GeV}$^{2}$, the red line is the newest fit including the 2009 RHIC data. Bottom panel: Truncated integral of $\Delta g$ for $0.001 \le x \le 0.05$ and $0.05 \le x \le 1$ at $Q^{2} = 10$ \textup{GeV}$^{2}$. Figure from \cite{DSSV_new}.}
\end{minipage}
\end{figure}

\section{Jet production at STAR to probe the gluon helicity distribution}
Next-to-leading-order (NLO) perturbative QCD (pQCD) can describe the inclusive jet and di-jet cross sections in proton-proton collisions at STAR \cite{Tai_thesis}. In 200 GeV proton-proton collisions, the mid-rapidity jet production is dominated by quark+gluon processes, whereas mid-rapidity jet production in 500 GeV proton-proton collisions are dominated by the quark+gluon and gluon+gluon processes. The double spin asymmetry $A_{LL}$ of jets/di-jets is proportional to the helicity distribution and spin asymmetry of polarized partons involved in the hard scattering. Therefore, the gluon polarization contribution to the proton can be directly probed by measuring the inclusive jet or di-jet $A_{LL}$. 

\subsection{Inclusive jet production at STAR}
In 2009 STAR recorded approximately 25 $pb^{-1}$ integrated luminosity of 200 GeV proton-proton collisions with the averaged polarization of beams around $55\%$. The relative uncertainty on the product of the polarization of the two beams is $6.5\%$. This data sample contains 20-fold more statistics than the published results in \cite{2006_jet_ALL}. Instead of the Mid-point cone jet algorithm applied in previous results, the anti-$k_{T}$ jet algorithm \cite{antikt} with R=0.6 is applied in the 2009 analysis to reduce the underlying event contribution and pile up backgrounds to the reconstructed jets. Jets reconstructed from the detector response (e.g. charged tracks or neutral energy depositions) in data are defined as detector level jets. Particle level jets which are based on final state particles and parton level jets which are formed from hard scattered partons (excluding those from the underlying events and beam remnants) can be reconstructed in the simulaton. Detector level jets can be reconstructed in simulation as well with the same trigger and detector conditions as in data. In simulation, a detector level jet and a parton or particle level jet are considered to be correlated when the distance between the two in $\eta-\varphi$ space is minimal and less than 0.5. The detector jet $p_{T}$ measured in data can be corrected to particle or parton jet $p_{T}$ according to the associations in simulation. The measured detector jet $A_{LL}$ values are corrected to parton jet $A_{LL}$ in each jet $p_{T}$ bin according to calculations in the simulation with polarized/unpolarized PDFs. Figure \ref{2009jet_plot} shows the 2009 STAR mid-rapidity inclusive jet double spin asymmetry $A_{LL}$ versus the parton jet $p_{T}$ in 200 GeV proton-proton collisions \cite{2009_jet_All}. The jet acceptance is divided into $|\eta|<0.5$ (top panel) and $0.5<|\eta|<1$ (bottom panel) to access different $x$ regions. Colored lines represent the NLO pQCD inclusive jet $A_{LL}$ calculations with different polarized parton distributions as inputs \cite{dis,dis1,theory1,theory2,theory5}. Data points fall between LSS10p \cite{dis} and DSSV \cite{theory1} model predictions.  

\begin{figure}[t]
\centerline{\includegraphics[width=30pc]{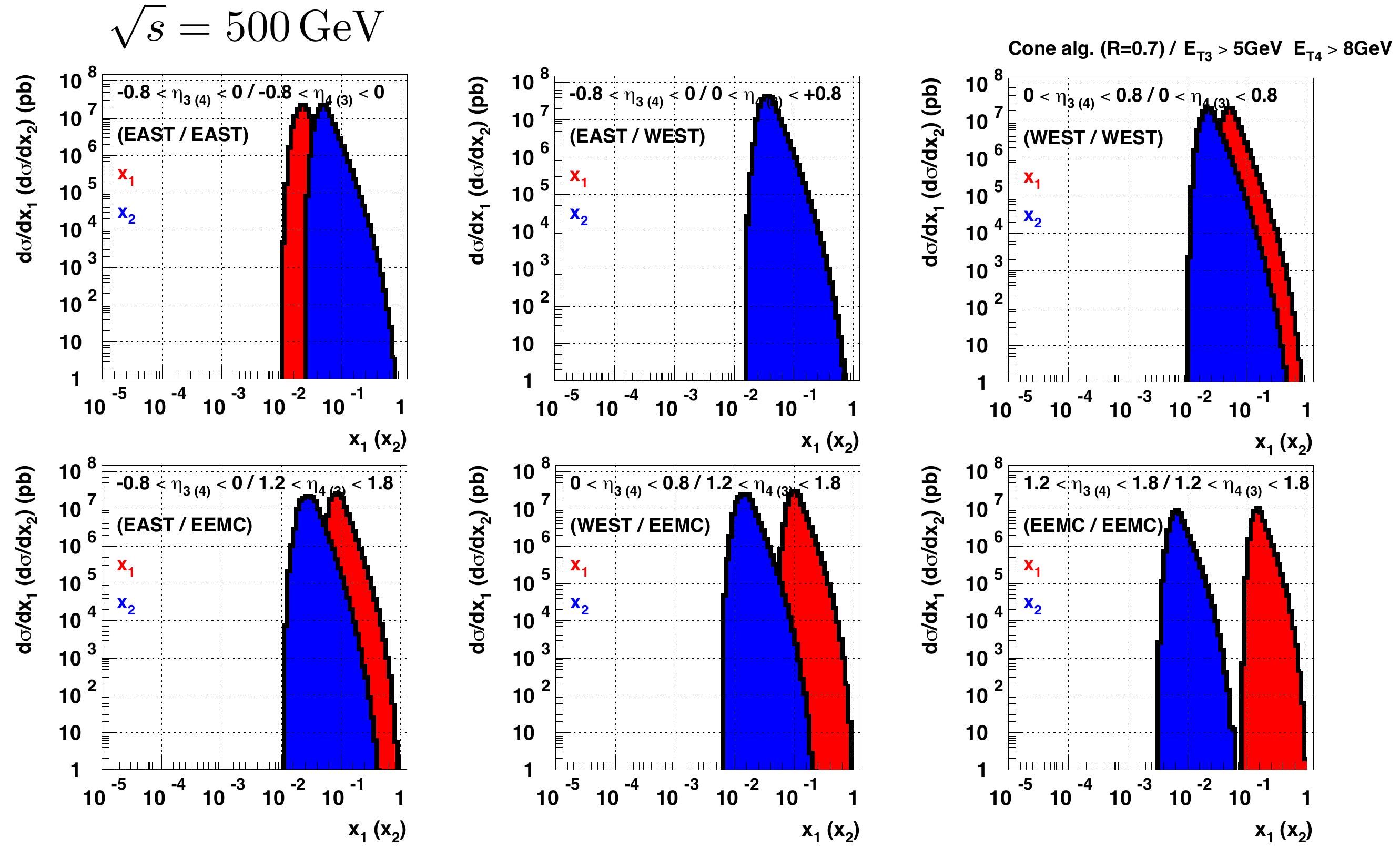}}
\caption{ \it $x_{1}$ (red) and $x_{2}$ (blue) of initial partons probed by di-jets reconstructed within the STAR detector acceptance ($-1<\eta<2$) in MC based on NLO thoery calculations. \cite{STAR_pp_pA_LOI}}
\label{fwd_central_x}
\end{figure}
\begin{figure}[t]
\centerline{\includegraphics[width=30pc]{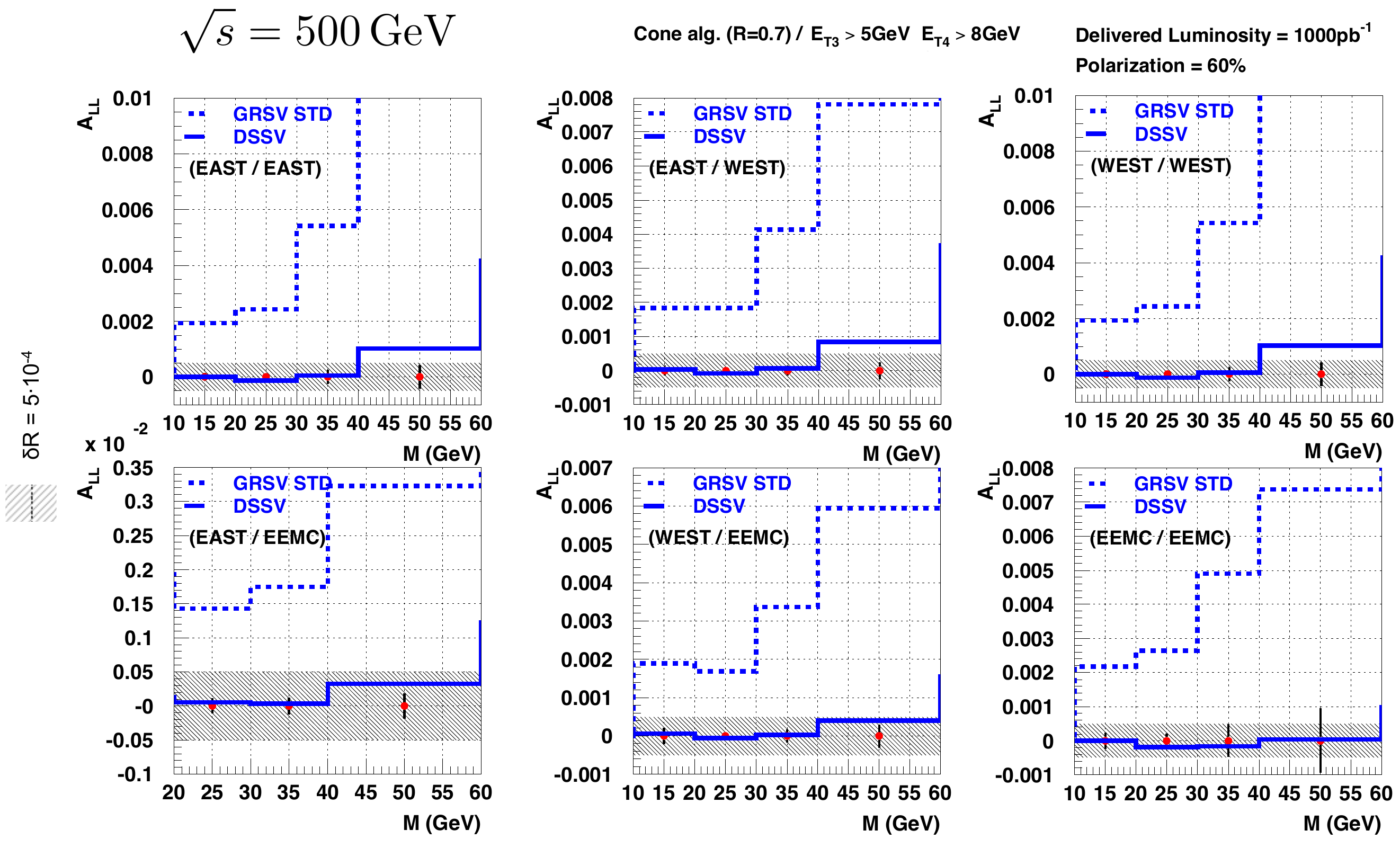}}
\caption{ \it Di-jet $A_{LL}$ versus invariant mass with different configurations in the STAR detector acceptance, $-1<\eta<2$ from MC based on NLO thoery calculations. \cite{STAR_pp_pA_LOI}}
\label{fwd_central_ALL}
\end{figure}

The NNPDF polarized PDFs are based on the DIS data, and a reweighting method is developed to include new experimental results \cite{dis1}. The DSSV group is the first to apply a global QCD fit on the combined data samples from DIS, Semi-Inclusive DIS (SIDIS) and RHIC pp experiments \cite{DSSV_new}. The latest DSSV update which includes the 2009 STAR inclusive jet $A_{LL}$ \cite{Pibero} and 2009 PHENIX inclusive $\pi^{0}$ $A_{LL}$ \cite{Phenix_pi0_ALL} exhibits a postive polarized gluon helicity function in the momentum fraction region $0.05<x<1$ as shown in the top panel of Figure \ref{DSSV_Dg}. The integral of the gluon helicity function in this $x$ region $\Delta G = \int_{0.05}^{1} \! \Delta g \, \mathrm{d}x (Q^{2} = 10$ GeV$^{2}) \approx 0.2$ is shown in the bottom panel of Figure \ref{DSSV_Dg} (red point within $90 \%$ confidence level (C.L.) represented by the blue band). A first non-zero gluon polarization contribution to the proton has been observed at RHIC.

RHIC is scheduled to operate another 200 GeV longitudianlly polarized proton-proton run in 2015. Future inclusive jet $A_{LL}$ measurements will provide better constraints with improved statistics on the gluon polarization in the same $x$ region \cite{2013_W_projection}. Dotted lines in the top panel of Figure \ref{DSSV_Dg} stand for alternative fits that are within $90\%$ confidence level limit, which reflects large uncertainties in derterming the gluon helicity function beyond the current $x$ range that can be accessed by RHIC. In additional to the 200 GeV longitudinal proton-proton run, RHIC also operated 500 GeV longitudinal proton-proton collisions with around $600 pb^{-1}$ delivered luminosity in 2012 and 2013. Gluon polarization in lower $x$ region will be probed by the ongoing 2012/2013 jet analysis in higher center of mass collision energy.

\subsection{Di-jet production at STAR}
Inclusive jets measured in 200 and 500 GeV proton-proton collisions can be used to probe the gluon polarization integral over a broad range in x. In the leading order QCD with collinear approximations, the momentum fractions of inital hard scattering partons  $x_{1}$ and $x_{2}$ are correlated with the di-jet invariant mass $M$ and the jet pseudo-rapidities $\eta_{3}$, $\eta_{4}$ ($M^{2} = x_{1}x_{2}s$, $ln(x_{1}/x_{2}) = \eta_{3}+\eta_{4}$). Therefore, a certain $x$ region can be selected by the di-jet production by varying the pseudorapidities of the jets. In the asymmetric partonic scattering, lower $x$ gluons can be probed by forward di-jet correlations. Uncertainties of the truncated moment of $\Delta g(x, Q^{2} = 10$ GeV$^{2})$ in the $0.001 \le x \le 0.05$ region (shown in the bottom panel of Figure \ref{DSSV_Dg}) remain large compared to the fitted results in the $0.05 \le x \le 1$ region. Achieving measurements which can probe this low $x$ region will help contribute to our understanding of the proton gluon spin distributions. STAR has nearly continuous, full azimuthal acceptance of electromagnetic calorimeters in $-1.0 < \eta < 4.0 $: the Barrel ElectroMagnetic Calorimeter (BEMC: $-1.0 < \eta < 1.0$), the End-cap ElectroMagnetic Calorimeter (EEMC: $1.07 < \eta < 2.0 $) and the Forward Meson Spectrometer (FMS: $2.3 < \eta < 4.0$). The electromagnetic calorimeters (and a possible forward hadronic calorimeter in the STAR upgrade plan \cite{STAR_pp_pA_LOI}) together with the Time Projection Chamber (TPC: $-1.3 < \eta < 1.3$) provide a broad $\Delta\eta\times\Delta\varphi$ coverage for di-jet measurements.

\begin{figure}[t]
\begin{minipage}{17pc}
\includegraphics[width=17pc]{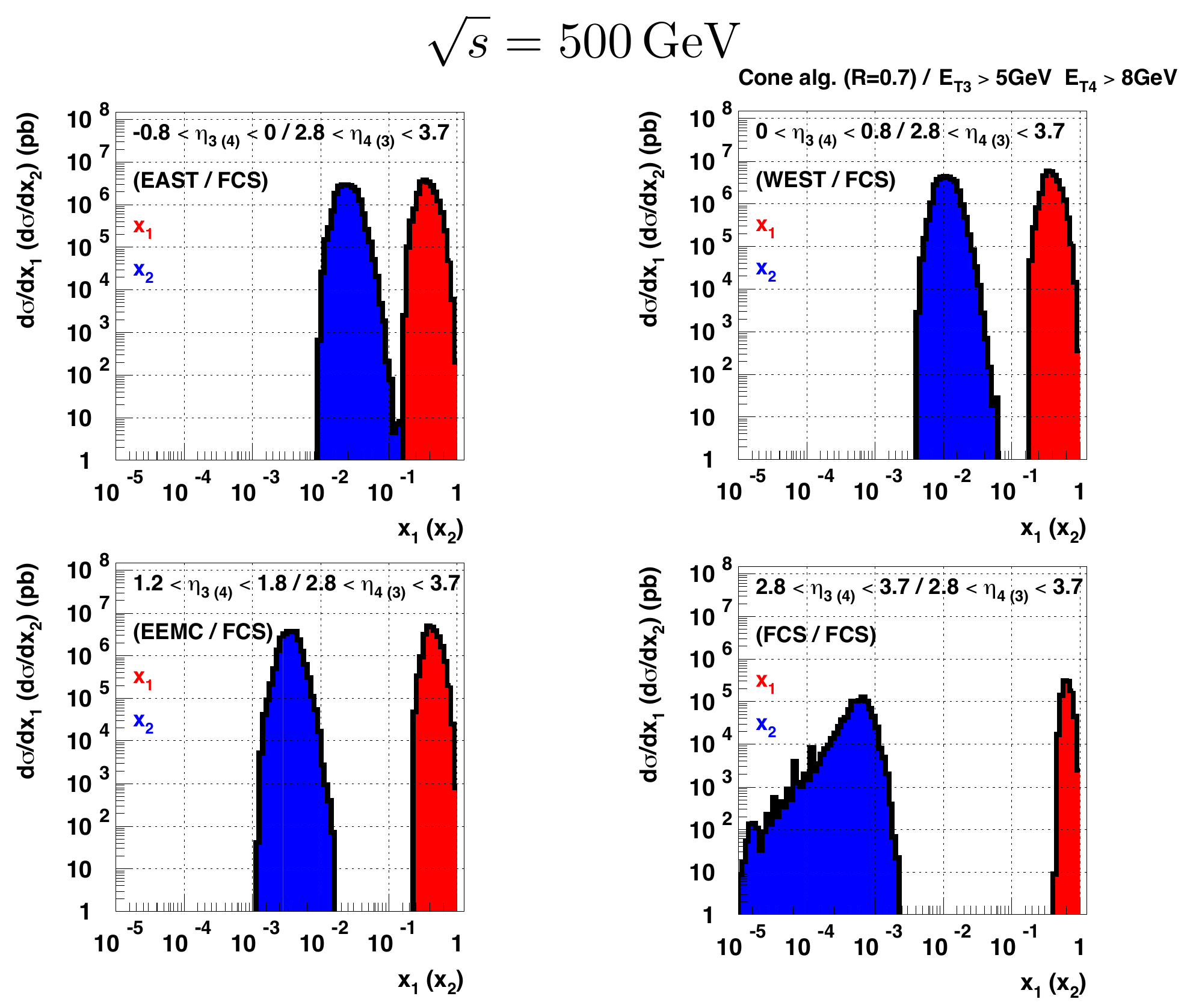} %\hspace{0.5pc}
\caption{\label{x1x2_fwd} \it $x_{1}$ (red) and $x_{2}$ (blue) of initial partons probed by forward di-jet with a proposed forward hadrnoinc calorimeter upgrade at STAR. \cite{STAR_pp_pA_LOI}}
\end{minipage} 
\hfill\begin{minipage}{20pc}
\includegraphics[width=20pc]{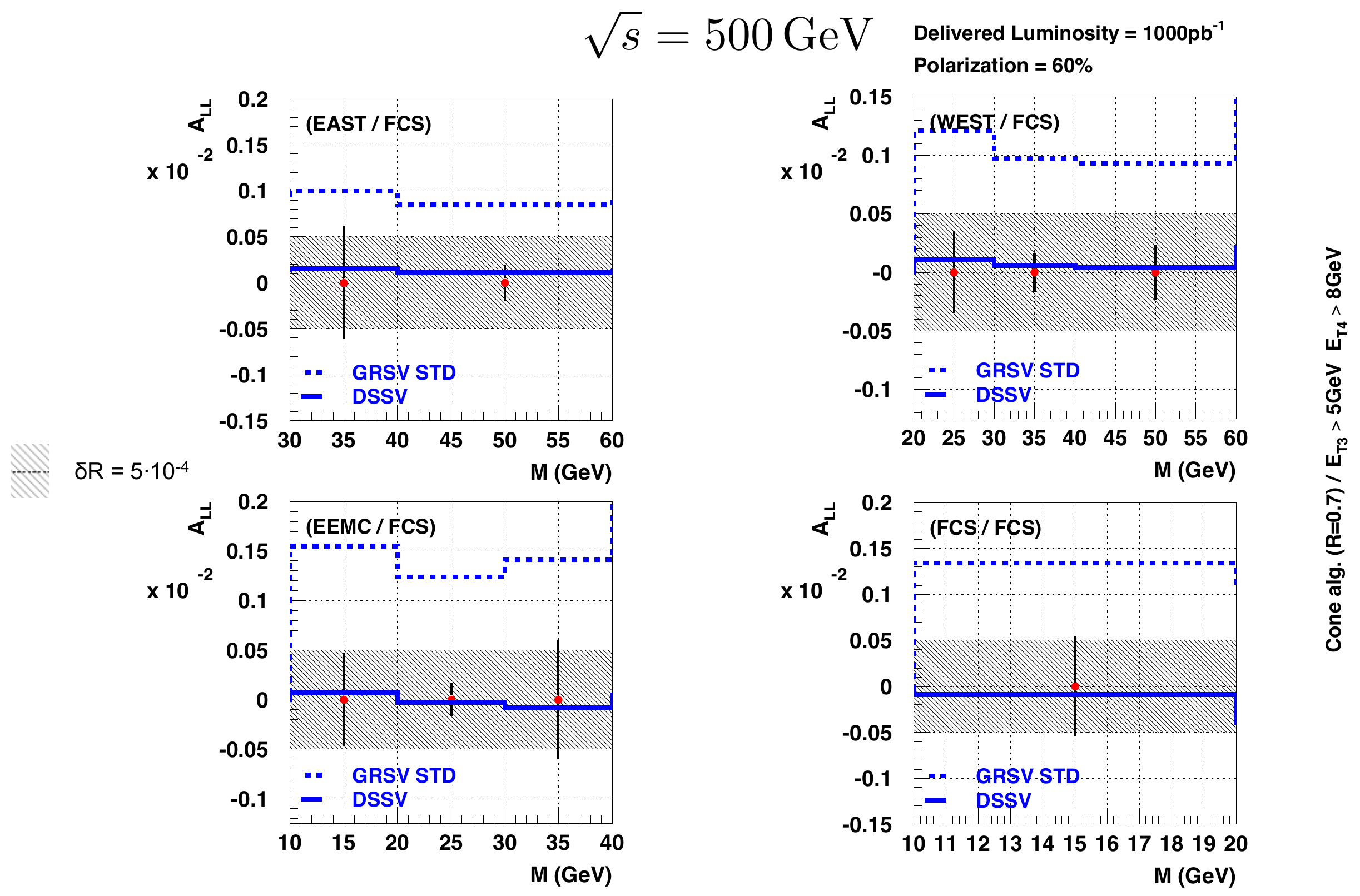}
\caption{\label{fwd_ALL} \it Projection of forward di-jet $A_{LL}$ asymmetries in 500 \textup{GeV} proton-proton collisions of 1000 $pb^{-1}$ delivered luminoisty with beam polarization of $60\%$. \cite{STAR_pp_pA_LOI}}
\end{minipage}
\end{figure}

Figure \ref{fwd_central_x} shows the measured intial parton $x_{1}$ and $x_{2}$ by final di-jets with six different detector configurations in 500 GeV proton-proton collisions from MC based on NLO QCD framework \cite{STAR_pp_pA_LOI}. The BEMC acceptance is divided into separate coverages: ``East" ($-1.0<\eta<0$) and ``West" ($0<\eta<1$). Jets are reconstructed using the Mid-point cone algorithm with cone radius R=0.7. In the di-jet pair, the leading jet $p_{T}$ is required to be larger than 8 $\textup{GeV/c}$ and the sub-leading jet $p_{T}$ should be larger than 5 $\textup{GeV/c}$. The current RHIC sensitive $x$ region extends to around $10^{-2}$ when both jets are reconstructed in the EEMC. The projected di-jet $A_{LL}$ for the West-West, West-East, East-East, West-EEMC, East-EEMC and EEMC-EEMC combinations are shown in Figure \ref{fwd_central_ALL} assuming the beam polarization is 60$\%$ and 1000 $pb^{-1}$ delivered luminosity \cite{STAR_pp_pA_LOI}. The dominant systematic error is the relative luminosity error $\delta R = 5\cdot10^{-4}$, the same value as cited in the 2009 inclusive jet $A_{LL}$ results. The statistical and systematic errors of the projected $A_{LL}$ as shown in Figure \ref{fwd_central_ALL} are much smaller than the separation between DSSV \cite{DSSV_new} and GRSV-STD \cite{GRSV} theory calculations, and both predict non-zero gluon helicity distribution. A forward hadronic calorimeter (FCS) is proposed in the STAR forward upgrade plan \cite{2013_W_projection}. Installation of this forward instrument will allow the di-jet measurements to go to the forward pseudorapidity and reach lower $x$ values. Projected initial parton $x_{1}$, $x_{2}$ values probed by forward di-jets detected in the FCS and associated detectors (West, East, EEMC, FCS) and their $A_{LL}$ asymmetries in NLO QCD MC are shown in Figure \ref{x1x2_fwd} and Figure \ref{fwd_ALL}. Fiducial volume cuts are applied as well for the forward di-jet simulation. The left-top panel is for the FCS and east Barrel di-jet production; the right-top is for the FCS and west Barrel di-jets; the left-bottom panel is for the FCS and EEMC di-jets and the right-bottom panel is for the FCS and FCS di-jets. The FCS-FCS di-jets probe the lowest $x$ region which is below $10^{-3}$. Reducing the systematic errors will be a primary goal to realize such measurements for studying gluon helicity distribution at low $x$ .

\section{W production at STAR to probe the sea quark helicity distribution}
The $\bar{u}$ and $\bar{d}$ momentum fraction distributions have been found asymmetric through Drell-Yan process by the E866 experiment \cite{Asym_sea_DY}. This phenomenon can not be fully described by perturbative QCD and indicates a non-perturbative mechanism may play a role in this field. In the parity violating weak processes, intial u (d) quark and $\bar{d}$ ($\bar{u}$ ) quark couple to $W^{+}$ ($W^{-}$) in proton-proton collisions. The longitudinally polarized 500 or 510 GeV proton-proton collisions at RHIC open a path to probe the light sea quark helicity functions via W productions ($p+p \rightarrow W^{+/-} + X \rightarrow l^{+/-} + X$). This process gets rid of the final state fragmentation functions that are encountered in semi-inclusive DIS processes. The W boson measured in high energy polarized proton-proton collisions is a unique probe for the polarized sea quark distrubtion at the W mass scale \cite{dis, sea_fit1}.
 
\begin{figure}[t]
\begin{minipage}{18pc}
\includegraphics[width=18pc]{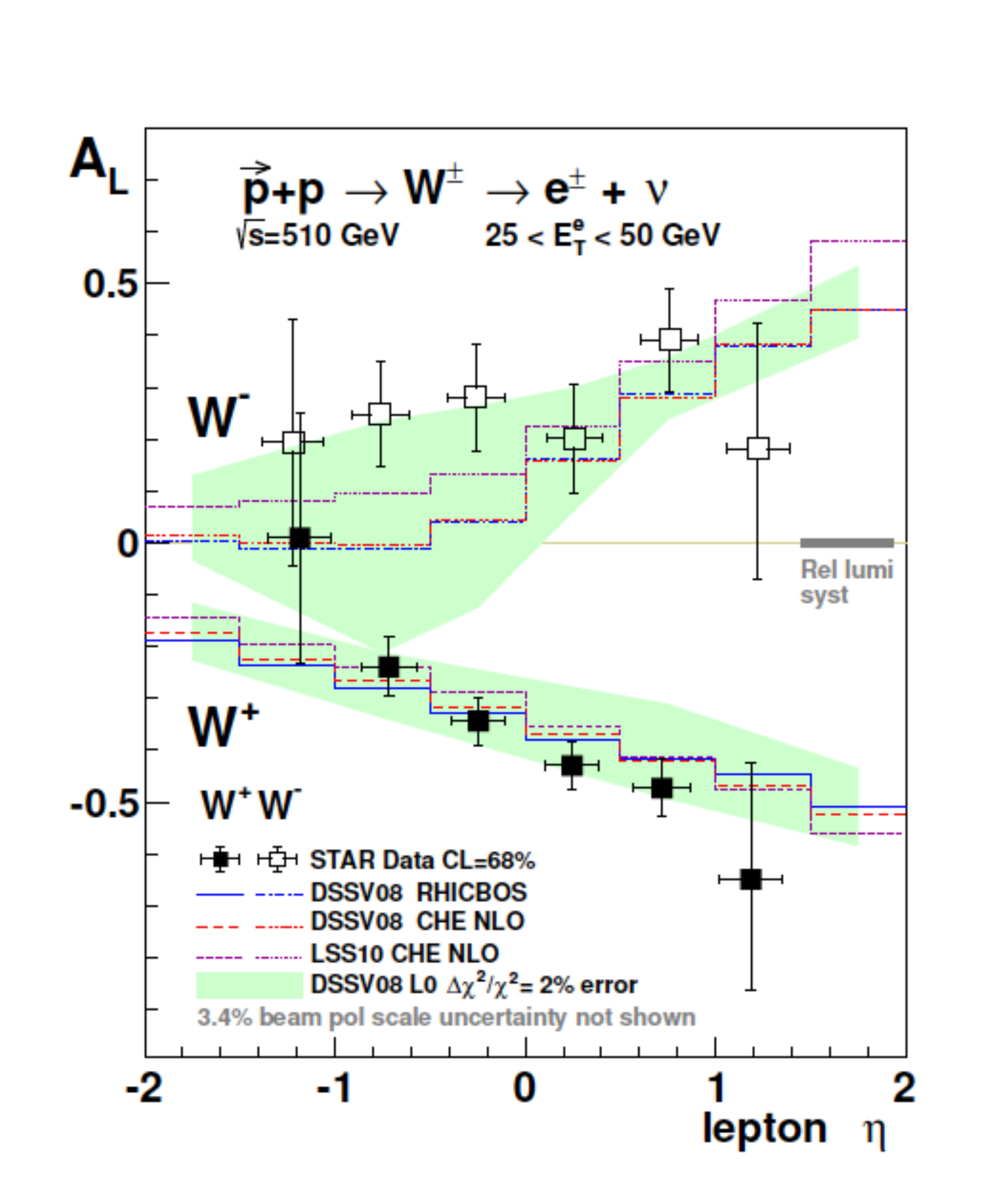} \hspace{2pc}
\caption{\label{2012_W_AL} \it 2011+2012 $W^{+/-}$ $A_{L}$ as a function of W decay lepton $\eta$ ($|\eta| < 1.3$) \cite{2012_W_AL_paper}. Color lines/band stand for theory predictions, see details in the text. The grey band presents the systematic error from the relative luminosity.}
\end{minipage} 
\hfill\begin{minipage}{17pc}
\includegraphics[width=17pc]{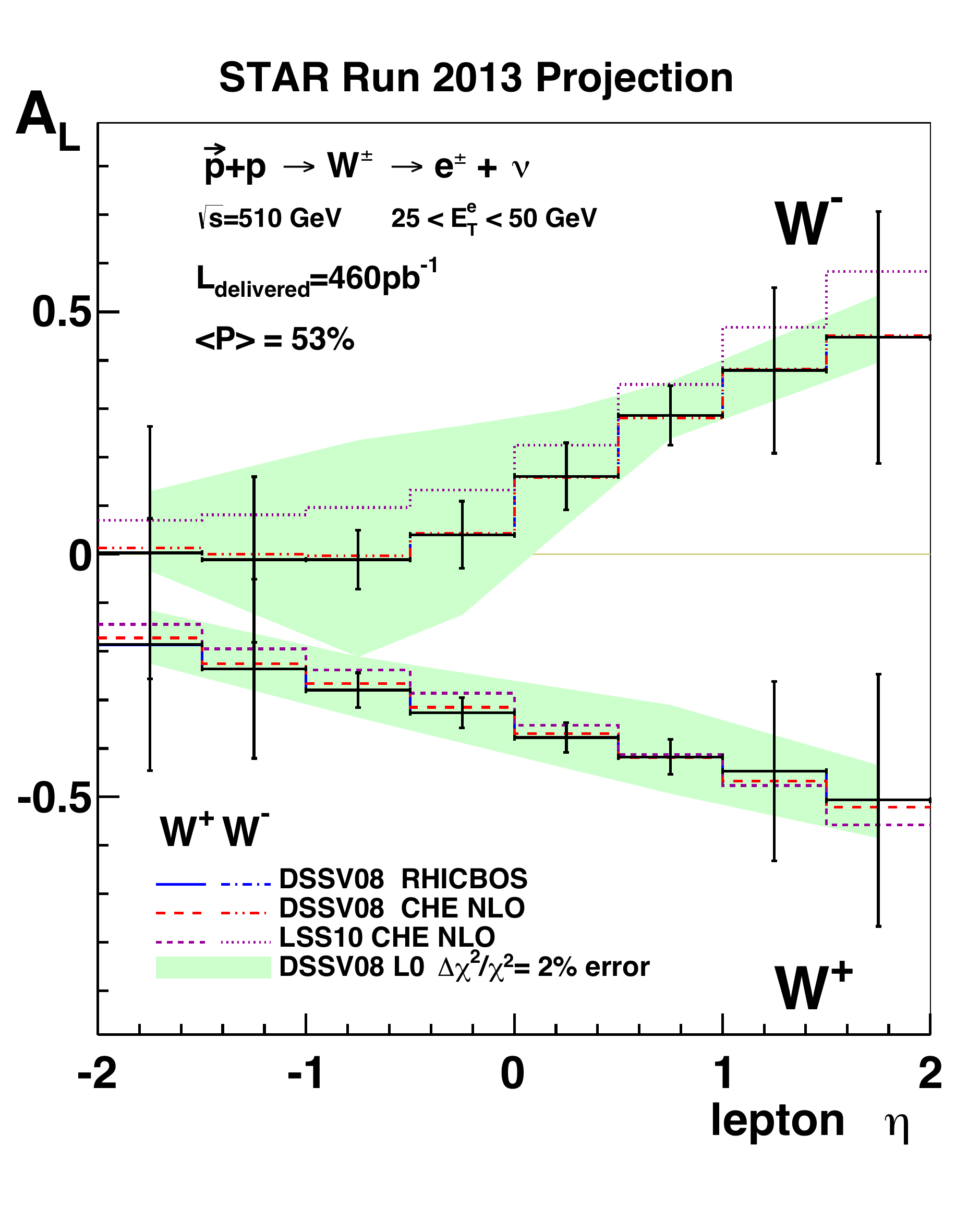}
\caption{\label{W_projection} \it Projections of $W^{+/-}$ $A_{L}$ in the 2013 510 \textup{GeV} p+p collisions at STAR with 460 $pb^{-1}$ integrated luminosity and average polarization $53\%$. Central values are evaluated by the CHE model with DSSV global QCD fit.}
\end{minipage}
\end{figure}

STAR has successfully reconstructed $W^{+}$/$W^{-}$ via their $e^{+}$/$e^{-}$ decay channels in the mid-rapidity ($-1<\eta<1$) from 2009 500 GeV proton-proton collisions \cite{2009_W_AL}. Measured mid-rapidity $W^{+/-}$ single spin asymmetries in 2009 data are consistent with predictions of the DSSV08 global QCD fit analysis which only include DIS and SIDIS results. But statistics were limited to separate different theoretical model descriptions. In RHIC runs of 2011 (500 GeV) and 2012 (510 GeV) longitudinally polarized proton-proton collisions, STAR collected around 94 $pb^{-1}$ integrated luminosity of data, with average proton beam polarizations of $49\%$ and $56\%$. The good statistics of this data sample allowed for the measurements of lepton pseudorapidity dependent W single spin asymmetry $A_{L}$. To obtain this goal, several updates were applied to the 2011 and 2012 combined data set. The detector acceptance to measure the W has been extended from $-1<\eta<1$ to $-1.3<\eta<1.3$, with the help from the shower maximum detector (SMD) of the EEMC and inner sectors of the TPC. The profile likehood method was used to combine W $A_{L}$ results which were measured independently in run 2011 and 2012 \cite{2012_W_AL_paper}. $W^{+/-}$ events are reconstructed from $W^{+/-}$ decays to $e^{+/-}$ with transverse energy $25<E_{T}^{e}<50$ (GeV). A significant signal to background ratio is seen in the reconstructed $W^{+/-}$ invariant mass distributions.

%As neutrinos can not be directly measured at STAR, a signed $p_{T}$ balance method is developed to reconstruced the W decay neutrino momentum. Backgrounds are evaluated from MC and the data driven QCD background is included.

\begin{figure}[t]
\centerline{\includegraphics[width=30pc]{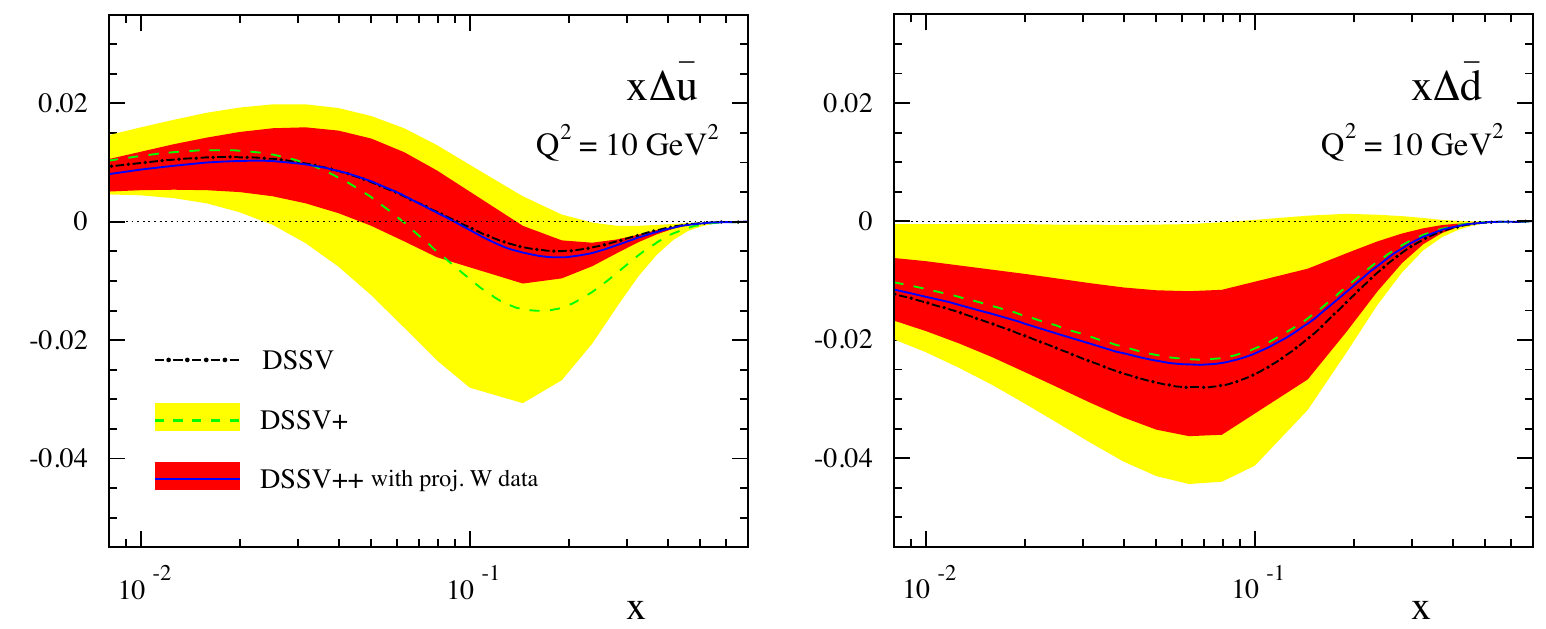}}
\caption{ \it $x$ dependent sea quark polarized PDF ($\Delta \bar{u}$ is shown in the left, and $\Delta \bar{d}$ is shown in the right) extracted from the DSSV global fit. Uncertainties change from the yellow band to the red bands after including the projection of combined 2009 to 2013 RHIC W $A_{L}$ asymmetries. Figure from \cite{2013_W_projection}.}
\label{Impact_sea_quark}
\end{figure}

The STAR W decay lepton pseudorapidity dependent single spin asymmetry $A_{L}$ from the 2011+2012 data is shown in Figure \ref{2012_W_AL}. Results of the $W^{+/-}$ $A_{L}$  are extracted from the likelihood function with $68\%$ confidence level. The grey band stands for the relative luminosity systematic error. Colored lines stand for the NLO QCD calculations on the $W^{+/-}$ $A_{L}$ in RHICBOS and CHE models with DSSV08/LSS10 \cite{dis, sea_fit1} helicty functions as inputs. The green band reflects the DSSV08 $\Delta \chi^{2} / \chi^{2} = 2\%$ uncertainties. This result probes the flavor separated light quark and anti-quark helicity distributions in the $0.05<x<0.2$ region. The $W^{+}$ $A_{L}$ which is consistent with the theory predictions indicates a negative $\bar{d}$ polarization. The $W^{-}$ $A_{L}$ which is sensitive to the $\bar{u}$ quark polarization is larger than the theory predictions for the $\eta_{e}<0$ range. This suggests a positive $\bar{u}$ helicity distribution. Therefore, a postive $\bar{u}-\bar{d}$ helcity dependent function is favored by this result.

The mid-rapidity $W^{+/-}$ $A_{L}$ asymmetries probe the polarization contributions carried by both valence quark and sea quark. Extending the pseudorapidity of the reconstructed $W^{+/-}$ to forward/backward regions will enhance the sensitivity for sea quark polarizations. A forward tracking system, the Forward GEM Tracker (FGT, $1<\eta<2$) was fully installed at STAR before the 2013 run. This update enhances the charge seperation capabilities of STAR for forward/backward electron/positrons. In 2013, RHIC delivered around 460 $pb^{-1}$ integrated luminosity 510 GeV proton-proton collisions which is the largest data sample so far. The tracking software with the FGT included is under development, and the FGT W analysis at STAR with the 2013 data is ongoing. The projection of lepton pseudorapidity dependent $W^{+/-}$ $A_{L}$ measured at STAR from the 2013 data is shown in Figure \ref{W_projection}. Central values of the projected W $A_{L}$ are based on the DSSV pQCD new fit which fits the STAR 2011+2012 W $A_{L}$ asymmetries. From 2009 to 2013, RHIC has delivered around 600 $pb^{-1}$ integrated luminosity 500/510 GeV longitudinally polarized proton-proton collisions. The extracted light sea quark polarized PDF $\Delta \bar{u}$ and $\Delta \bar{d}$ from the DSSV analysis including the combined 2009 to 2013 W $A_{L}$ projection from STAR and PHENIX are shown in Figure \ref{Impact_sea_quark} \cite{2013_W_projection}. The yellow bands represent the uncertainties constrained by the DIS data. Uncertainties of the helicity dependent PDFs reduce significantly after including the combined RHIC $W^{+/-}$ $A_{L}$ projection as shown in the red bands of Figure \ref{Impact_sea_quark}. The RHIC W program with the combined 2009 to 2013 data will improve our understanding of the light sea quark polarization contributions to the proton spin structure.

\section{Summary and Outlook}
Latest measurements of the STAR longitudinal progam improve our understanding of the proton spin structure. A non-zero gluon spin contribution in the range of $0.05<x<1$ is indicated from the 2009 inclusive jet double spin asymmetry $A_{LL}$ results. The total gluon spin contribution might be at the same order of the quark polarization contribution to the proton; this needs future measurements to prove. The W decay lepton pseudorapidity dependent $W^{+/-}$ single spin asymmetries provide constraints on the $\bar{u}$ and $\bar{d}$ polarized PDFs at the W mass scale. A postive $\bar{u}$ polarized PDF is suggested based on the STAR data and theoretical model comparisons in $0.05<x<0.2$ kinematic region.

Uncertainties of both polarized gluon and sea quark PDFs at low $x$ remain large compared to the currently probed region. Longer 500 GeV longitudinally polarized proton-proton operations at RHIC and possible forward upgrade will provide opportunities to measure di-jet and W productions at STAR in the forward pseudorapidity with high precison data. Such observables are under study and future analysis will impact the dermination of the gluon, $\bar{u}$ and $\bar{d}$ helicity dependent PDFs and provide new inputs for the proton spin consistuation studies in the extended $x$ regions.

\section{References}
\bibliographystyle{iopart-num}
\bibliography{xuanli}
\end{document}